\documentclass[aps,prl,reprint,showpacs,amsmath,amssymb,superscriptaddress,floatfix]{revtex4-1}
\usepackage{graphicx}
\usepackage{ifpdf}
\usepackage{epstopdf}
\begin{document}

\title{Universal Duality in Luttinger Liquid Coupled to Generic Environment}

\author{Igor V.\ Yurkevich}
\affiliation{Nonlinearity and Complexity Research Group, Aston University, Birmingham B4 7ET, United Kingdom}

\author{Oleg M.\ Yevtushenko}
\affiliation{Ludwig Maximilians University, Arnold Sommerfeld Center and Center for Nano-Science, Munich, DE-80333, Germany}

\begin{abstract}
We study a Luttinger Liquid (LL) coupled to a generic environment consisting of bosonic modes with arbitrary
density-density and current-current interactions. The LL can be either in the conducting phase and perturbed
by a weak scatterer or in the insulating phase and perturbed by a weak link. The environment modes can
also be scattered by the imperfection in the system with arbitrary transmission and reflection amplitudes.
We present a general method of calculating correlation functions under the presence of the environment
and prove the duality of exponents describing the scaling of the weak scatterer and of the weak link. This
duality holds true for a broad class of models and is sensitive to neither interaction nor environmental modes
details, thus it shows up as the universal property. It
ensures that the environment cannot generate new stable fixed points of the RG flow. Thus, the LL always
flows toward either conducting or insulating phase. Phases are separated by a sharp
boundary which is shifted by the influence of the environment. Our results are relevant, for example, for low-energy transport
in (i) an interacting quantum wire or a carbon nanotube where the electrons are coupled to the acoustic phonons scattered by
the lattice defect; (ii)~a mixture of interacting fermionic and bosonic cold atoms where the bosonic
modes are scattered due to an abrupt local change of the interaction, (iii) mesoscopic electric circuits.
\end{abstract}

\date{\today}


\maketitle

The Luttinger Liquid (LL) is the canonical model which describes low energy properties of low dimensional
interacting systems \cite{Tom:50,Lutt:63,HALDANE:81}. Its applicability is amazingly broad (see the book
[\onlinecite{Giamarchi}] for a review) and ranges from quantum wires and carbon nanotubes \cite{CN-LL}, to
Fractional- and  Spin- Quantum Hall systems \cite{FQH-1,FQH-2,SQH-1,SQH-2}, to mesoscopic
electric circuits \cite{SafiSaleur,KindermannNazarov}, to name a few. Recent increasing interest in the LLs
has been stimulated by the theoretical progress in understanding physics of cold gases \cite{ColdGasesRev1}
and  topological insulators \cite{Shen} where the LL again plays the role of one the basic models
\cite{ColdGasesRev2,AAYu}. Quasiparticles in the LL are collective waves, or plasmons, described
by the universal low energy theory which can be derived for fermionic \cite{GogNersTsv} and for bosonic
\cite{CazalillaBos} interacting systems. It yields a power-law decay of correlation functions which
have been detected experimentally via conductance measurements and a scanning tunneling microscopy
both in carbon nanotubes \cite{Yao:99,Ishii:03,Lee:04} and quantum nanowires
\cite{Auslaender:02,Slot:04,Levy:06,Kim:06}. The correlation functions, in their turn, govern scaling
behavior of different perturbations, applied to the LL, and related physical observables. For example,
low energy transport in the LL is extremely sensitive to imperfections. Two archetypal cases are usually
considered: (i)~the ideal LL, i.e., perfectly conducting phase (CP), can be perturbed by a short range
weak scatterer (WS); (ii)~two ideal LLs on disjoint left/right half-axes, i.e., insulating  phase (IP), can be
connected via  a weak tunneling contact -- the weak link  (WL) perturbation. Both perturbations scale
with changing the smallest energy in the system, which can be temperature or bias. Their scaling
exponents, $ {\Delta}_{\rm ws}$ and ${\Delta}_{\rm wl} $, are system dependent
\cite{KaneFis:92a, KF:92b, MatYueGlaz:93, FurusakiNagaosa:93b, Furusaki:97,EggertAffleck:95, FabrizioGogolin:95},
however, they are related by the universal formula:
\begin{equation}\label{Duality}
  {\Delta}_{\rm ws} \times {\Delta}_{\rm wl}  = 1 \, .
\end{equation}
Eq.(\ref{Duality}) is often referred to as the {\it duality relation} between the WS and the WL. It was shown
\cite{KaneFis:92a} that Eq.(\ref{Duality}) follows from the duality of fields whose correlation functions
yield $ \Delta_{\rm ws} $ and $ \Delta_{\rm wl} $. The perturbation is relevant (irrelevant)
if its scaling dimension is smaller (greater) than Euclidean space dimension $ d $; $ d = 1 $ for a local perturbation. Therefore,
the duality (\ref{Duality}) asserts that only one of these two perturbations is relevant, the second one being irrelevant.
Reformulating in the RG language, only one phase, either the CP or the IP, corresponds to the stable fixed point
of the RG flow, the second phase is unstable. This immediately explains transport properties of the system:
if the WS is irrelevant and the WL is relevant the system is driven to the CP and it is driven to the IP in the
opposite case. The duality relation has been demonstrated for a single LL not coupled to anything else, for 
example surrounding bath or environment. Without environment attractive fermions and repulsive bosons fall 
into the CP limit while repulsive fermions - to the IP one.

The natural question is whether the duality is robust and survives in more complex systems, like the LL coupled
via the density-density interaction to a single massless bosonic mode,  e.g. to the mode of acoustic phonons
\cite{LL-Ph,GYL:2011,GYL:2011a}. Surprisingly, elaborated and rather lengthy calculations have shown that the
duality (\ref{Duality}) holds true for arbitrary set of systems parameters, including the reflection amplitude
of the phonon mode from the imperfection \cite{YuGYeL}. It has been assumed that the phonon sees the
imperfection as an elastic point-like scatterer. Unfortunately, the cumbersome calculus did not allow the authors
to identify the origin of the duality. The breakthrough has been achieved by one of us in the recent paper \cite{YuDual}.
By using a new method, $ \, N \, $ coupled LLs have been considered for the case of the generic inter-mode interactions
which may include density-density and/or current-current channels. For $ \, N_c \, $  conducting-
and $ \, N_i = N - N_c \, $  insulating phases a universal description of the
Green's functions has been obtained which has manifested the
duality of two sets of the local correlation functions under interchanging CP and IP subsystems. This explains the origin
of the duality  discovered in Ref.\cite{YuGYeL} in the cases of freely propagating- and fully reflected phonons. In spite
of a noticeable progress, only several particular cases were considered in \cite{YuGYeL,YuDual} and the answers were
obtained after lengthy algebra. For instance, only fully-reflected or freely propagating scatterers were addressed in Ref.\cite{YuGYeL};
transport at small but finite energy would not allow the classification of the subsystems onto the CPs and the IPs and
it is beyond the scope of Ref.\cite{YuDual}.
Therefore, a complete explanation of the duality origin and a clear understanding of its applicability were still missing.

\begin{figure}[t]
{\includegraphics[width=0.85\columnwidth,clip]{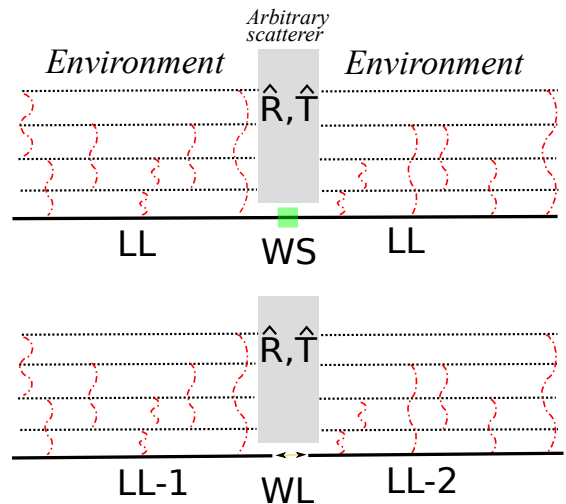}}
\caption{\label{Cartoon}
(colored on-line)
The system under consideration: the environment consists of arbitrary number of bosonic
channels (dotted lines) which interact (red wavy lines) with each other and with the Luttinger
Liquid, the inter-channel interaction is also implied though not drawn. The bosonic modes
can be scattered by the imperfection (shadowed box) described
by the arbitrary scattering/transmission amplitudes, $ \hat{R}, \ \hat{T} $. The LL is either
in the conducting phase (upper panel) and is perturbed by a weak scatterer (green box);
or it is in the insulating phase (lower panel) and is perturbed by a weak tunneling (arrows
connecting two half-axes).
}
\end{figure}

In this Letter, we present an improved version of the analytical method which allows us to remove all
 listed above restrictions of the previous studies and to clearly formulate the class of systems where the duality of
the given scaling exponents is present. The system which we consider consists of the LL coupled to $ \, N - 1 $
massless modes of the environment, see a sketch on Fig.\ref{Cartoon}. The LL is in one of the
phases, either the CP, the upper panel of  Fig.\ref{Cartoon}, or the IP, the lower panel of  Fig.\ref{Cartoon}.
The intra- and inter-channel interactions are arbitrary: density-density and/or the current-current interactions between
all modes and channels. The current and the density are
proportional to the derivatives of the chiral bosonic fields and, therefore, the effective bosonized action is quadratic.
The imperfection, located at the point $ \, x = 0 $, is capable of driving the LL from the WS
to the WL setup and it may cause an arbitrary single particle scattering of the environmental modes. The only assumption
on the environmental modes scattering is that they can be described within quadratic theory in the bosonic degrees of freedom with
proper boundary conditions at $ \, x = 0 $. Particles forming LL are assumed to be of a different nature from those
providing environmental bath. Therefore, local scatterer enables transitions within LL and within the environment but
not mixing these two subspaces. Without loss of generality,  we assume the inversion (left$\leftrightarrow$right) symmetry
of translational invariant system and the reflection symmetry when a scatterer is placed at the origin.

We rigorously prove that the duality relation (\ref{Duality}) holds true under these very general
conditions. This, in particular, means that the environment is capable of neither modifying the stability of the RG fixed points
for the single LL, nor creating the new ones, unless multiparticle scattering is taken into account
or the RG flow becomes multidimensional due to additional nonlinearities. Note that, in this Letter, we consider
only a single LL coupled to quadratic bath and postpone above extensions of the model for further studies.

Let us now specify the model in more details and outline main steps of the calculations. Detailed
derivations can be found in the Supplemental Materials. We assume that $N$ 1D-channels each supporting two chiral modes
(labeled by $\eta={\rm R,\,L}$) are described by the Lagrangian density
\begin{align}\label{L0}
   L=\frac{1}{4\pi}{\boldsymbol\Theta}^{\rm T} \,\left({\hat \tau}_3\,\partial_t+{\hat V}\,\partial_x\right)\,\partial_x {\boldsymbol \Theta} \,,
\end{align}
with bosonic real fields
\begin{align}
{\boldsymbol\Theta}^{\rm T} = \left({\boldsymbol\theta}_{\rm R}^{\rm T},{\boldsymbol\theta}_{\rm L}^{\rm T}\right)\,,\,\,
{\boldsymbol\theta}^{\rm T}_{\eta}=\left(\theta_{\eta}^{(1)}, \ldots, \theta_{\eta}^{(N)}\right)\, ;
\end{align}
and with real time running over both Keldysh branches. Pauli matrices $ {\hat \tau}_j \, $  act in chiral space, i.e. in space
of right- left-movers. The densities of chiral modes in $i$th channel are related to the chiral fields as $ \, \rho^{(i)}_{\rm R/L}
= \pm\,\partial_x \theta^{(i)}_{\rm R/L}/2\pi $. Diagonal entries of the $2N\times 2N$ matrix $ \, {\hat V} \, $ describe
chiral channels with velocities being renormalized due to the intra-channel interactions; the off-diagonal elements
are strengths of inter-channel interactions between chiral densities. The matrix $ \, {\hat V} \, $
is obviously real and symmetric but its exact structure is not important for our purposes.
We only assume the presence of inversion symmetry, i.e. the Lagrangian is invariant under transformation
\begin{equation}\label{Invers}
{\boldsymbol\Theta}(x) \to{\hat\tau_1}\,{\boldsymbol\Theta}(-x) \, .
\end{equation}
Note that 1) the choice of origin is irrelevant so far since we are dealing with the translational invariant system at the moment;
and 2) that inversion symmetry implies that arbitrary density-density interaction between chiral channels is reduced to 
density-density and current-current channel non-chiral interactions. Thus, the symmetries of the matrix ${\hat V}$ are

\begin{align}\label{symV}
{\hat V}={\hat V}^{\rm T} \, , \qquad {\hat\tau_1}\,{\hat V}\,{\hat\tau_1}={\hat V} \, .
\end{align}

Let us now add the imperfection at $ x = 0 $. It can be naturally described by implying the boundary (matching) conditions.
We choose to use the transfer-matrix~$ {\hat {\cal T}} $
\begin{align}\label{BoundCond-T}
    {\boldsymbol\Theta}\left(+0\right)={\hat{\cal T}} \,{\boldsymbol\Theta}\left(-0\right) \, .
\end{align}
The transfer-matrix $ {\hat {\cal T}} $ must respect reflection symmetry, i.e. inversion symmetry around the origin, under 
the transformation (\ref{Invers}),
which requires
\begin{align}\label{BoundCond-2}
    {\hat\tau_1} \, {\hat{\cal T}} \, {\hat\tau_1} = {\hat{\cal T}}^{-1} \, .
\end{align}
We use notations for the fields ${\boldsymbol\theta}_{\eta}(\pm 0)={\boldsymbol\theta}_{\eta}(x\!=\!\pm 0,\omega)$
in the frequency domain, the latter allows one to accommodate possible energy dependence of scattering amplitudes.
The transfer-matrix can be written in terms of  entries of the scattering matrix  $ \, {\hat S} $
\begin{align}\label{BoundCond-S}
\left(
                                \begin{array}{c}
                                  {\boldsymbol\theta}_{\rm R}(+0) \\
                                  {\boldsymbol\theta}_{\rm L}(-0) \\
                                \end{array}
                              \right)\,= {\hat S}
\left(
                                \begin{array}{c}
                                  {\boldsymbol\theta}_{\rm L}(+0) \\
                                  {\boldsymbol\theta}_{\rm R}(-0) \\
                                \end{array}
                              \right) ,
\end{align}
which also must respect reflection symmetry and, therefore, obeys ${\hat\tau_1}\,{\hat S}\,{\hat\tau_1} = {\hat S}$
and can be parametrized with only two $N\times N$ reflection and transmission matrices:
\begin{equation}\label{S-matr-Symm}
     {\hat S} =
                       \left(
                           \begin{array}{cc}
                                   {\hat R} & {\hat T} \\
                                   {\hat T} & {\hat R} \\
                           \end{array}
                       \right) \, .
\end{equation}
The transfer-matrix is expressed in terms of $ {\hat T} \mbox{ and } {\hat R} \, $
as follows:
\begin{align}\label{TS}
{\hat{\cal T}}=\left(
           \begin{array}{cc}
             {\hat T} - {\hat R} {\hat T}^{-1} {\hat R} & {\hat R} {\hat T}^{-1} \\
             -{\hat T}^{-1} {\hat R} & {\hat T}^{-1}  \\
           \end{array}
         \right)\,.
\end{align}
In the model which we consider the particles forming LL (the first channel) cannot be transformed into environment
particles at the impurity since they all are of different nature; therefore, $ \, \hat{R} \mbox{ and } \hat{T} \, $ are 
block diagonal:
\begin{equation}\label{R-T-matr}
  \hat{R} =
   \left(
      \begin{array}{cc}
         R & 0 \\
             0    & \hat{r}
      \end{array}
   \right), \
  \hat{T} =
   \left(
      \begin{array}{cc}
         T & 0 \\
             0    & \hat{t}
      \end{array}
   \right).
\end{equation}
Here, the LL channel and the environment are described by scalars $ \, R \mbox{ and } T \, $ and by  $(N-1)\times (N-1)$
matrices $ \, \hat{r} \mbox{ and }  \hat{t} $, respectively.

Let us diagonalize the Lagrangian by a transformation
\begin{eqnarray}\label{RotFields}
{\boldsymbol\Theta}(x) = {\hat{\cal M}}\,{\tilde{\boldsymbol\Theta}}(x)
\end{eqnarray}
which must: 1) keep $ \, {\hat \tau}_3 \, $ invariant
\begin{equation}\label{Symmetry1}
       \hat{\cal M}^{\rm T} \hat{\tau}_3 \hat{\cal M} = \hat{\tau}_3
\end{equation}
[cf. the first term in Eq.(\ref{L0})]; and 2) simultaneously diagonalize the matrix $ \, {\hat V} \, $ which
contains interactions
\begin{align}\label{V}
  {\hat v}={\hat {\cal M}}^{\rm T}\,{\hat V}\,{\hat {\cal M}}={\rm diag}(v_1,\dots,v_N; v_1,\ldots, v_N) \, .
\end{align}
The matrix $\hat{\cal M} $ is real and belongs to the pseudo-orthogonal group $ O( N, N ) $, see Eq.(\ref{Symmetry1}).
We note that Eq.(\ref{Symmetry1}) is equivalent to the preservation of the Kac-Moody algebra in operator technique:
\begin{eqnarray}
 [ \hat{\boldsymbol\Theta}(x), \hat{\boldsymbol\Theta}^{\rm T}(x')] = i \pi \, {\rm sign}(x-x')\, \hat\tau_3 \, ,
\label{K-M}
\end{eqnarray}
with $ [{\hat {\boldsymbol\Theta}}(x), {\hat {\boldsymbol\Theta}}^{\rm T}(x')] \equiv {\hat{\boldsymbol\Theta}}(x) \otimes
{\hat{\boldsymbol\Theta}}^{\rm T}(x') - {\hat{\boldsymbol\Theta}}(x') \otimes {\hat{\boldsymbol\Theta}}^{\rm T}(x)$.

In Eq. (\ref{V}), we have taken into account the reflection symmetry on the initial (interacting) problem via the
relation $ \, {\hat v} = {\hat \tau}_1 \, {\hat v} \, {\hat \tau}_1 $ which holds true if
\begin{equation}\label{Symmetry2}
       \hat{\tau}_1\, \hat{\cal M}\, \hat{\tau}_1= \hat{\cal M} \, .
\end{equation}
Eqs. (\ref{Symmetry1}) and (\ref{Symmetry2}) define the full symmetry of the transformation matrix ${\hat{\cal M}}$.

To resolve all above formulated symmetry restrictions, we can parametrize the transformation matrix ${\hat{\cal M}}$ 
with a single $N\times N$ matrix ${\hat M}$ :
\begin{align}\label{M-parametr}
{\hat{\cal M}}=\frac{1-\tau_1}{2}\,\otimes\,{\hat M}+\frac{1+\tau_1}{2}\,\otimes\, \left({\hat M}^{-1}\right)^{\rm T}\,.
\end{align}

After diagonalization of the Lagrangian, one obtains the theory in terms of free non-interacting
fields $ \tilde{\boldsymbol\Theta} $ scattered by the imperfection.
These free fields have a very simple Green's function $ i {\hat{\tilde{G}}} = < \!\! {\tilde{\boldsymbol\Theta}}
\otimes {\tilde{\boldsymbol\Theta}}^{\rm T} \!\!\! > $ which can be written in the standard scattering state
representation \cite{Suppl-1}.
The latter requires boundary
conditions for $ \tilde{\boldsymbol\Theta} $ which are obtained after inserting Eq.(\ref{RotFields})
into Eq.(\ref{BoundCond-T}):
\begin{align}\label{t-m}
   {\tilde {\boldsymbol\Theta}}(+0)={\hat{\tilde{\cal T}}}\,{\tilde{\boldsymbol\Theta}}(-0) \, , \
         {\hat{\tilde{\cal T}}}={\hat {\cal M}}^{-1} \, {\hat{\cal T}}\,{\hat {\cal M}} \, .
\end{align}
Due to the symmetry (\ref{Symmetry2}), the new transfer matrix ${\hat{\tilde{\cal T}}}$ has the same structure
as ${\hat{\cal T}}$, cf. Eq.(\ref{TS}), but with new scattering amplitudes, $({\hat{\tilde R}}, \ {\hat{\tilde T}})$ \cite{Suppl-2}.

The Green's function of the original fields, $ {\hat G} $, can be found from the transformation (\ref{RotFields}):
\begin{align}\label{RotGF}
   {\hat G}(x,x') = {\hat{\cal M}}\,{\hat {\tilde G}}(x,x')\,{\hat{\cal M}}^{\rm T} \, .
\end{align}
The matrix $  {\hat{\cal M}} $ depends neither on coordinates nor on time. Therefore,
the relation between $ {\hat G} $ and $ {\hat {\tilde G}} $ is local.
Eq.(\ref{RotGF}) completes the formal description of the system so we have all information
which is necessary to calculate the scaling dimension of the WS and of the WL.

These perturbations acting in the LL-channel are described by \cite{YuDual}:
\begin{align}
\label{pert-1}
    L_{\rm ws} &= \lambda\, \cos \Bigl( \Phi(T=1) \Bigr) \, , \\
\label{pert-2}
    L_{\rm wl} &= t\, \cos \Bigl( \Phi (T=0) \Bigr) \,,
\end{align}
where the field $ \, \Phi \, $ is the difference between two (right- and left-) incoming chiral fields
of the 1st channel
\begin{align}\label{in}
    \Phi(T)=\theta^{(1)}_{\rm R} (x=-0, t)-\theta^{(1)}_{\rm L} (x=+0, t)\,.
\end{align}
The notation in Eqs. (\ref{pert-1}-\ref{in}) stresses that the field $ \, \Phi \, $ (and its correlation function) depends on 
boundary conditions and, in particular, on boundary conditions in the LL-channel. In Eq. (\ref{pert-1}) we assume 
that the LL-channel is in the CP with $ \, T = 1, \, R = 0 \, $ and it is perturbed by $L_{\rm ws}$. In the IP with 
$ \, T = 0, \, R = 1 \, $ the LL-channel is perturbed by $L_{\rm wl}$. After integrating out high-energy degrees of
freedom with the energy lying above the running cut-off $ \, \varepsilon$, the Green's function ${\cal G}=
-i\langle \Phi \, \Phi \rangle$ with retarded component
$$
{\cal G}=-\frac{2\pi i}{\omega + i 0}\,\Delta_{11}(T)
$$
defines one-loop RG equations
$$
\frac{\partial\ln\lambda}{\partial\ln\varepsilon}=\Delta_{\rm ws}-1\,,\quad
\frac{\partial\ln t}{\partial\ln\varepsilon}=\Delta_{\rm wl}-1 \, ;
$$
where 
\begin{equation}\label{ScExps}
   \Delta_{\rm ws}=\Delta_{11}(T=1) \, , \
   \Delta_{\rm wl}=\Delta_{11}(T=0) \, .
\end{equation}
The scattering and the tunneling amplitudes acquire an
effective power-low dependence on the smallest energy scale:
\begin{equation}
   \lambda (\varepsilon) \sim \lambda\, \varepsilon^{\Delta_{\rm ws} - 1} , \quad
   t  (\varepsilon) \sim t\, \varepsilon^{\Delta_{\rm wl} - 1} .
\end{equation}
The scaling exponent $\Delta_{11}(T)$ is found from the local Green's function at the origin and
can be written as "11"-element of the $ \, N \times N \, $ matrix $ \, {\hat\Delta} \, $ defined in the channel space:
\begin{eqnarray}\label{HatDelta}
{\hat\Delta}(T) & = & \frac{1}{2}{\hat M}\,\left[1-{\hat{\tilde R}}+{\hat{\tilde T}}\right]\,{\hat M}^{\rm T}\\\nonumber
 & + & \frac{1}{2}\left({\hat M}^{-1}\right)^{\rm T}\,\left[1+{\hat{\tilde R}}-{\hat{\tilde T}}\right]\,{\hat M}^{-1}\,.
\end{eqnarray}
Eq.(\ref{HatDelta}) can be reduced to \cite{Suppl-3}:
\begin{align}\label{Delta-answ}
{\hat\Delta}(T)=\left[{\hat\xi}^{-1}+{\hat\delta}^{-1}\right]^{-1}
+\left[{\hat\xi}+{\hat\delta}\right]^{-1} \, ,
\end{align}
where
\begin{align}\label{xi}
{\hat\xi}=\frac{1-({\hat R}-{\hat T})}{1+({\hat R}-{\hat T})}\,,\quad
{\hat\delta}={\hat M}\,{\hat M}^{\rm T}\,.
\end{align}
The matrix ${\hat\delta}$ depends on interaction only and defines the scaling exponent in the absence of environment.
Taking into account the block-diagonal structure of the matrices $ \, \hat{R} \mbox{ and } \hat{T} $, Eq.(\ref{R-T-matr}),
the matrix $ \, {\hat\xi} \, $ is also block diagonal
\begin{equation}\label{xi-matr}
{\hat\xi}=\left(
            \begin{array}{cc}
              \xi & 0 \\
              0 & {\hat\zeta} \\
            \end{array}
          \right)
\end{equation}
where we have introduced the scalar $ \, \xi \, $ and the $ \, (N-1) \times (N-1) \, $ matrix $ \, {\hat \zeta} $
\begin{align}
\xi=\frac{1-(R-T)}{1+(R-T)}\,,\quad {\hat\zeta}=\frac{1-({\hat r}-{\hat t})}{1+({\hat r}-{\hat t})}\,.
\end{align}
The scalar $\xi$ describes the phase of the LL: 1) $ \, \xi=\infty \, $ for conducting phase
($T=1, R=0$)  perturbed by WS; and 2) $ \, \xi=0 \, $  for insulating phase ($T=0, R=1$)
perturbed by WL. The matrix ${\hat\zeta}$ describes scattering of bath modes. It is arbitrary
because no additional assumptions were implied, i.e. by the proper choice of the reflection
and the transmission matrices we can take into account the arbitrary environmental scattering.

Finally, combining Eqs.(\ref{ScExps}-\ref{xi-matr}) we arrive at our main result:
\begin{align}\label{ws-f}
   \Delta_{\rm ws} &=\lim_{\xi\to\infty}{\hat\Delta}_{11}= \left[ {\hat\delta}^{-1}+\left(
                                                 \begin{array}{cc}
                                                   0 & 0 \\
                                                   0 & {\hat{\zeta}^{-1}} \\
                                                 \end{array}
                                               \right)
  \right]^{-1}_{11} \,,\\\label{wl-f}
   \Delta_{\rm wl} &=\lim_{\xi\to 0}{\hat\Delta}_{11}= \left[ {\hat\delta}+ \left(
                                                 \begin{array}{cc}
                                                   0 & 0 \\
                                                   0 & {\hat{\zeta}} \\
                                                 \end{array}
                                               \right)\right]^{-1}_{11} \,.
\end{align}
It gives explicit expressions for scaling dimensions of operators which  perturb the LL. The duality relation
Eq.(\ref{Duality}) directly results from the structure of matrices in Eqs.(\ref{ws-f},\ref{wl-f}) \cite{Suppl-4}.
The duality is universal: details of bath scattering, ${\hat{\zeta}}$, and of interactions, ${\hat\delta}$, are
all {\it irrelevant}.


To conclude, we have proven that the duality between the weak-scatterer- and the weak-link scaling
exponents holds true in the system where the Luttinger Liquid is coupled to a very generic environment.
The duality guarantees the same classification of the fixed stable points of the RG flow as in the isolated LL
\cite{KaneFis:92a, KF:92b}, the coupling to the environment is unable to create new stable fixed points.
The duality and the structure of the RG flow are universal and very robust because (i) the duality relation
is insensitive to parameters of the LL,  of the environment, and of the coupling; (ii)~the type of the coupling does
not influence the duality, so it can be arbitrary chiral density--density interactions; (iii)~the scattering of the
environmental modes by the imperfection can be arbitrary, the duality and the RG fixed points survive even
opening the system when the number of coherent particles at the imperfection is not conserved (but cf.
\cite{AGR}). We have restricted ourselves to a
system with the reflection symmetry. This assumption is only a technical simplification which can be easily removed
without changing our results. The necessary condition for the duality of the scaling exponents is the Kac-Moody
algebra, Eq.(\ref{K-M}) [or related duality of linear combinations of the chiral fields $ \, {\boldsymbol \theta}_{\rm R}
\pm {\boldsymbol \theta}_{\rm L} $]. It leads to the symmetry relation Eq.(\ref{Symmetry1}) and allows one
to derive Eq.(\ref{Delta-answ}). Based on our results, we can define the universality class where the duality is always 
present due to the symmetry Eq.(\ref{Symmetry1}): it includes the systems where there are no:
1)~direct transitions between the LL and the environment at the imperfection; 2)~ multiparticle scattering; 
3)~additional nonlinearities except those which are related to the WS- and the WL perturbations. Presence 
of any of these three effects could change the RG drastically but such extensions are beyond the scope of 
the present Letter and will be considered elsewhere.

In real experiments, our model is applicable, for instance, to an interacting quantum wire or a carbon nanotube
where the electrons are coupled to the acoustic phonons scattered by the lattice defect \cite{LL-Ph}.  Another
physical realization is a mixture of interacting fermionic and bosonic cold atoms where the
bosonic modes are scattered due to an abrupt local change of the interaction~\cite{InhomLL}.

\begin{acknowledgments}
O.M.Ye. acknowledges support from the DFG through SFB TR-12, and the Cluster of Excellence, Nanosystems
Initiative Munich. I.V.Yu. acknowledges hospitality of the Ludwig Maximilians University, Arnold Sommerfeld Center,
Munich.

\end{acknowledgments}

\bibliography{my}

\begin{thebibliography}{10}%
\makeatletter
\providecommand \@ifxundefined [1]{%
 \ifx #1\undefined \expandafter \@firstoftwo
 \else \expandafter \@secondoftwo
\fi
}%
\providecommand \@ifnum [1]{%
 \ifnum #1\expandafter \@firstoftwo
 \else \expandafter \@secondoftwo
\fi
}%
\providecommand \enquote [1]{``#1''}%
\providecommand \bibnamefont  [1]{#1}%
\providecommand \bibfnamefont [1]{#1}%
\providecommand \citenamefont [1]{#1}%
\providecommand\href[0]{\@sanitize\@href}%
\providecommand\@href[1]{\endgroup\@@startlink{#1}\endgroup\@@href}%
\providecommand\@@href[1]{#1\@@endlink}%
\providecommand \@sanitize [0]{\begingroup\catcode`\&12\catcode`\#12\relax}%
\@ifxundefined \pdfoutput {\@firstoftwo}{%
 \@ifnum{\z@=\pdfoutput}{\@firstoftwo}{\@secondoftwo}%
}{%
 \providecommand\@@startlink[1]{\leavevmode\special{html:<a href="#1">}}%
 \providecommand\@@endlink[0]{\special{html:</a>}}%
}{%
 \providecommand\@@startlink[1]{%
  \leavevmode
  \pdfstartlink
   attr{/Border[0 0 1 ]/H/I/C[0 1 1]}%
   user{/Subtype/Link/A<</Type/Action/S/URI/URI(#1)>>}%
  \relax
 }%
 \providecommand\@@endlink[0]{\pdfendlink}%
}%
\providecommand \url  [0]{\begingroup\@sanitize \@url }%
\providecommand \@url [1]{\endgroup\@href {#1}{\urlprefix}}%
\providecommand \urlprefix [0]{URL }%
\providecommand \Eprint[0]{\href }%
\@ifxundefined \urlstyle {%
  \providecommand \doi [1]{doi:\discretionary{}{}{}#1}%
}{%
  \providecommand \doi [0]{doi:\discretionary{}{}{}\begingroup
  \urlstyle{rm}\Url }%
}%
\providecommand \doibase [0]{http://dx.doi.org/}%
\providecommand \Doi[1]{\href{\doibase#1}}%
\providecommand \bibAnnote [3]{%
  \BibitemShut{#1}%
  \begin{quotation}\noindent
    \textsc{Key:}\ #2\\\textsc{Annotation:}\ #3%
  \end{quotation}%
}%
\providecommand \bibAnnoteFile [2]{%
  \IfFileExists{#2}{\bibAnnote {#1} {#2} {\input{#2}}}{}%
}%
\providecommand \typeout [0]{\immediate \write \m@ne }%
\providecommand \selectlanguage [0]{\@gobble}%
\providecommand \bibinfo [0]{\@secondoftwo}%
\providecommand \bibfield [0]{\@secondoftwo}%
\providecommand \translation [1]{[#1]}%
\providecommand \BibitemOpen[0]{}%
\providecommand \bibitemStop [0]{}%
\providecommand \bibitemNoStop [0]{.\EOS\space}%
\providecommand \EOS [0]{\spacefactor3000\relax}%
\providecommand \BibitemShut [1]{\csname bibitem#1\endcsname}%
\bibitem{Tom:50}%
  \BibitemOpen
  \bibfield{author}{%
  \bibinfo {author} {\bibfnamefont{S.}~\bibnamefont{Tomonaga}},\ }%
  \bibfield{journal}{%
  \bibinfo {journal} {Prog. Theor. Phys.}\ }%
  \textbf{\bibinfo {volume} {5}},\ \bibinfo {pages} {544} (\bibinfo {year}
  {1950})%
  \bibAnnoteFile{NoStop}{Tom:50}%
\bibitem{Lutt:63}%
  \BibitemOpen
  \bibfield{author}{%
  \bibinfo {author} {\bibfnamefont{J.~M.}\ \bibnamefont{Luttinger}},\ }%
  \bibfield{journal}{%
  \bibinfo {journal} {J. Math. Phys.}\ }%
  \textbf{\bibinfo {volume} {4}},\ \bibinfo {pages} {1154} (\bibinfo {year}
  {1963})%
  \bibAnnoteFile{NoStop}{Lutt:63}%
\bibitem{HALDANE:81}%
  \BibitemOpen
  \bibfield{author}{%
  \bibinfo {author} {\bibfnamefont{F.~D.~M.}\ \bibnamefont{Haldane}},\ }%
  \bibfield{journal}{%
  \bibinfo {journal} {J. Phys. {\rm C}}\ }%
  \textbf{\bibinfo {volume} {14}},\ \bibinfo {pages} {2585} (\bibinfo {year}
  {1981})%
  \bibAnnoteFile{NoStop}{HALDANE:81}%
\bibitem{Giamarchi}%
  \BibitemOpen
  \bibfield{author}{%
  \bibinfo {author} {\bibfnamefont{T.}~\bibnamefont{Giamarchi}},\ }%
  \emph{\bibinfo {title} {Quantum Physics in One Dimension}}\ (\bibinfo
  {publisher} {Cla\-ren\-don Press},\ \bibinfo {address} {London},\ \bibinfo
  {year} {2004})%
  \bibAnnoteFile{NoStop}{Giamarchi}%
\bibitem{CN-LL}%
  \BibitemOpen
  \bibfield{author}{%
  \bibinfo {author} {\bibfnamefont{M.}~\bibnamefont{Bockrath}}, \bibinfo
  {author} {\bibfnamefont{D.~H.}\ \bibnamefont{Cobden}}, \bibinfo {author}
  {\bibfnamefont{J.}~\bibnamefont{Lu}}, \bibinfo {author}
  {\bibfnamefont{A.~G.}\ \bibnamefont{Rinzler}}, \bibinfo {author}
  {\bibfnamefont{R.~E.}\ \bibnamefont{Smalley}}, \bibinfo {author}
  {\bibfnamefont{L.}~\bibnamefont{Balents}},\ and\ \bibinfo {author}
  {\bibfnamefont{P.~L.}\ \bibnamefont{McEuen}},\ }%
  \bibfield{journal}{%
  \bibinfo {journal} {Nature}\ }%
  \textbf{\bibinfo {volume} {397}},\ \bibinfo {pages} {598} (\bibinfo {year}
  {1999})%
  \bibAnnoteFile{NoStop}{CN-LL}%
\bibitem{FQH-1}%
  \BibitemOpen
  \bibfield{author}{%
  \bibinfo {author} {\bibfnamefont{C.~L.}\ \bibnamefont{Kane}}\ and\ \bibinfo
  {author} {\bibfnamefont{M.~P.~A.}\ \bibnamefont{Fisher}},\ }%
  \bibfield{journal}{%
  \bibinfo {journal} {Phys. Rev. {\rm B}}\ }%
  \textbf{\bibinfo {volume} {51}},\ \bibinfo {pages} {13449} (\bibinfo {year}
  {1995})%
  \bibAnnoteFile{NoStop}{FQH-1}%
\bibitem{FQH-2}%
  \BibitemOpen
  \bibfield{author}{%
  \bibinfo {author} {\bibfnamefont{C.~L.}\ \bibnamefont{Kane}}, \bibinfo
  {author} {\bibfnamefont{R.}~\bibnamefont{Mukhopadhyay}},\ and\ \bibinfo
  {author} {\bibfnamefont{T.~C.}\ \bibnamefont{Lubensky}},\ }%
  \bibfield{journal}{%
  \bibinfo {journal} {Phys. Rev. Lett.}\ }%
  \textbf{\bibinfo {volume} {88}},\ \bibinfo {pages} {036401} (\bibinfo {year}
  {2002})%
  \bibAnnoteFile{NoStop}{FQH-2}%
\bibitem{SQH-1}%
  \BibitemOpen
  \bibfield{author}{%
  \bibinfo {author} {\bibfnamefont{C.~Y.}\ \bibnamefont{Hou}}, \bibinfo
  {author} {\bibfnamefont{E.}~\bibnamefont{Kim}},\ and\ \bibinfo {author}
  {\bibfnamefont{C.}~\bibnamefont{Chamon}},\ }%
  \bibfield{journal}{%
  \bibinfo {journal} {Phys. Rev. Lett.}\ }%
  \textbf{\bibinfo {volume} {102}},\ \bibinfo {pages} {076602} (\bibinfo {year}
  {2009})%
  \bibAnnoteFile{NoStop}{SQH-1}%
\bibitem{SQH-2}%
  \BibitemOpen
  \bibfield{author}{%
  \bibinfo {author} {\bibfnamefont{J.~C.~Y.}\ \bibnamefont{Teo}}\ and\ \bibinfo
  {author} {\bibfnamefont{C.~L.}\ \bibnamefont{Kane}},\ }%
  \bibfield{journal}{%
  \bibinfo {journal} {Phys. Rev. {\rm B}}\ }%
  \textbf{\bibinfo {volume} {79}},\ \bibinfo {pages} {235321} (\bibinfo {year}
  {2009})%
  \bibAnnoteFile{NoStop}{SQH-2}%
\bibitem{SafiSaleur}%
  \BibitemOpen
  \bibfield{author}{%
  \bibinfo {author} {\bibfnamefont{I.}~\bibnamefont{Safi}}\ and\ \bibinfo
  {author} {\bibfnamefont{H.}~\bibnamefont{Saleur}},\ }%
  \bibfield{journal}{%
  \bibinfo {journal} {Phys. Rev. Lett.}\ }%
  \textbf{\bibinfo {volume} {93}},\ \bibinfo {pages} {126602} (\bibinfo {year}
  {2004})%
  \bibAnnoteFile{NoStop}{SafiSaleur}%
\bibitem{KindermannNazarov}%
  \BibitemOpen
  \bibfield{author}{%
  \bibinfo {author} {\bibfnamefont{M.}~\bibnamefont{Kindermann}}\ and\ \bibinfo
  {author} {\bibfnamefont{Y.~V.}\ \bibnamefont{Nazarov}},\ }%
  \bibfield{journal}{%
  \bibinfo {journal} {Phys. Rev. Lett.}\ }%
  \textbf{\bibinfo {volume} {91}},\ \bibinfo {pages} {136802} (\bibinfo {year}
  {2003})%
  \bibAnnoteFile{NoStop}{KindermannNazarov}%
\bibitem{ColdGasesRev1}%
  \BibitemOpen
  \bibfield{author}{%
  \bibinfo {author} {\bibfnamefont{I.}~\bibnamefont{Bloch}}, \bibinfo {author}
  {\bibfnamefont{J.}~\bibnamefont{Dalibard}},\ and\ \bibinfo {author}
  {\bibfnamefont{W.}~\bibnamefont{Zwerger}},\ }%
  \bibfield{journal}{%
  \bibinfo {journal} {Rev. Mod. Phys.}\ }%
  \textbf{\bibinfo {volume} {80}},\ \bibinfo {pages} {885 } (\bibinfo {year}
  {2008})%
  \bibAnnoteFile{NoStop}{ColdGasesRev1}%
\bibitem{Shen}%
  \BibitemOpen
  \bibfield{author}{%
  \bibinfo {author} {\bibfnamefont{S.~Q.}\ \bibnamefont{Shen}},\ }%
  \emph{\bibinfo {title} {Topological Insulators}}\ (\bibinfo {publisher}
  {Springer},\ \bibinfo {address} {Springer Berlin Heidelberg},\ \bibinfo
  {year} {2012})%
  \bibAnnoteFile{NoStop}{Shen}%
\bibitem{ColdGasesRev2}%
  \BibitemOpen
  \bibfield{author}{%
  \bibinfo {author} {\bibfnamefont{M.~A.}\ \bibnamefont{Cazalilla}}, \bibinfo
  {author} {\bibfnamefont{R.}~\bibnamefont{Citro}}, \bibinfo {author}
  {\bibfnamefont{T.}~\bibnamefont{Giamarchi}}, \bibinfo {author}
  {\bibfnamefont{E.}~\bibnamefont{Orignac}},\ and\ \bibinfo {author}
  {\bibfnamefont{M.}~\bibnamefont{Rigol}},\ }%
  \bibfield{journal}{%
  \bibinfo {journal} {Rev. Mod. Phys.}\ }%
  \textbf{\bibinfo {volume} {83}},\ \bibinfo {pages} {1405 } (\bibinfo {year}
  {2011})%
  \bibAnnoteFile{NoStop}{ColdGasesRev2}%
\bibitem{AAYu}%
  \BibitemOpen
  \bibfield{author}{%
  \bibinfo {author} {\bibfnamefont{B.~L.}\ \bibnamefont{Altshuler}}, \bibinfo
  {author} {\bibfnamefont{I.~L.}\ \bibnamefont{Aleiner}},\ and\ \bibinfo
  {author} {\bibfnamefont{V.~I.}\ \bibnamefont{Yudson}},\ }%
  \bibfield{journal}{%
  \bibinfo {journal} {Phys. Rev. Lett.}\ }%
  \textbf{\bibinfo {volume} {111}},\ \bibinfo {pages} {086401} (\bibinfo {year}
  {2013})%
  \bibAnnoteFile{NoStop}{AAYu}%
\bibitem{GogNersTsv}%
  \BibitemOpen
  \bibfield{author}{%
  \bibinfo {author} {\bibfnamefont{A.~O.}\ \bibnamefont{Gogolin}}, \bibinfo
  {author} {\bibfnamefont{A.~A.}\ \bibnamefont{Nersesyan}},\ and\ \bibinfo
  {author} {\bibfnamefont{T.~A.}\ \bibnamefont{M}},\ }%
  \emph{\bibinfo {title} {Bosonization and Strongly Correlated Systems}}\
  (\bibinfo {publisher} {Cambridge University Press},\ \bibinfo {address}
  {Cambridge},\ \bibinfo {year} {2004})%
  \bibAnnoteFile{NoStop}{GogNersTsv}%
\bibitem{CazalillaBos}%
  \BibitemOpen
  \bibfield{author}{%
  \bibinfo {author} {\bibfnamefont{M.~A.}\ \bibnamefont{Cazalilla}},\ }%
  \bibfield{journal}{%
  \bibinfo {journal} {Journal of Physics B: AMOP}\ }%
  \textbf{\bibinfo {volume} {37}},\ \bibinfo {pages} {S1 } (\bibinfo {year}
  {2004})%
  \bibAnnoteFile{NoStop}{CazalillaBos}%
\bibitem{Yao:99}%
  \BibitemOpen
  \bibfield{author}{%
  \bibinfo {author} {\bibfnamefont{Z.}~\bibnamefont{Yao}}, \bibinfo {author}
  {\bibfnamefont{H.~W.~C.}\ \bibnamefont{Postma}}, \bibinfo {author}
  {\bibfnamefont{L.}~\bibnamefont{Balents}},\ and\ \bibinfo {author}
  {\bibfnamefont{C.}~\bibnamefont{Dekker}},\ }%
  \bibfield{journal}{%
  \bibinfo {journal} {Nature}\ }%
  \textbf{\bibinfo {volume} {402}},\ \bibinfo {pages} {273} (\bibinfo {month}
  {NOV 18}\ \bibinfo {year} {1999})%
  \bibAnnoteFile{NoStop}{Yao:99}%
\bibitem{Ishii:03}%
  \BibitemOpen
  \bibfield{author}{%
  \bibinfo {author} {\bibfnamefont{H.}~\bibnamefont{Ishii}} \emph{et~al.},\ }%
  \bibfield{journal}{%
  \bibinfo {journal} {Nature}\ }%
  \textbf{\bibinfo {volume} {426}},\ \bibinfo {pages} {540} (\bibinfo {year}
  {2003})%
  \bibAnnoteFile{NoStop}{Ishii:03}%
\bibitem{Lee:04}%
  \BibitemOpen
  \bibfield{author}{%
  \bibinfo {author} {\bibfnamefont{J.}~\bibnamefont{Lee}}, \bibinfo {author}
  {\bibfnamefont{S.}~\bibnamefont{Eggert}}, \bibinfo {author}
  {\bibfnamefont{H.}~\bibnamefont{Kim}}, \bibinfo {author}
  {\bibfnamefont{S.~J.}\ \bibnamefont{Kahng}}, \bibinfo {author}
  {\bibfnamefont{H.}~\bibnamefont{Shinohara}},\ and\ \bibinfo {author}
  {\bibfnamefont{Y.}~\bibnamefont{Kuk}},\ }%
  \bibfield{journal}{%
  \bibinfo {journal} {Phys. Rev. Lett.}\ }%
  \textbf{\bibinfo {volume} {93}},\ \bibinfo {pages} {166403} (\bibinfo {year}
  {2004})%
  \bibAnnoteFile{NoStop}{Lee:04}%
\bibitem{Auslaender:02}%
  \BibitemOpen
  \bibfield{author}{%
  \bibinfo {author} {\bibfnamefont{O.~M.}\ \bibnamefont{Auslaender}}, \bibinfo
  {author} {\bibfnamefont{A.}~\bibnamefont{Yacoby}}, \bibinfo {author}
  {\bibfnamefont{R.}~\bibnamefont{de~Picciotto}}, \bibinfo {author}
  {\bibfnamefont{K.~W.}\ \bibnamefont{Baldwin}}, \bibinfo {author}
  {\bibfnamefont{L.~N.}\ \bibnamefont{Pfeiffer}},\ and\ \bibinfo {author}
  {\bibfnamefont{K.~W.}\ \bibnamefont{West}},\ }%
  \bibfield{journal}{%
  \bibinfo {journal} {Science}\ }%
  \textbf{\bibinfo {volume} {295}},\ \bibinfo {pages} {825} (\bibinfo {year}
  {2002})%
  \bibAnnoteFile{NoStop}{Auslaender:02}%
\bibitem{Slot:04}%
  \BibitemOpen
  \bibfield{author}{%
  \bibinfo {author} {\bibfnamefont{E.}~\bibnamefont{Slot}}, \bibinfo {author}
  {\bibfnamefont{M.~A.}\ \bibnamefont{Holst}}, \bibinfo {author}
  {\bibfnamefont{H.~S.~J.}\ \bibnamefont{van~der Zant}},\ and\ \bibinfo
  {author} {\bibfnamefont{S.~V.}\ \bibnamefont{Zaitsev-Zotov}},\ }%
  \bibfield{journal}{%
  \Doi{10.1103/PhysRevLett.93.176602}{\bibinfo {journal} {Phys. Rev. Lett.}}\
  }%
  \textbf{\bibinfo {volume} {93}},\ \bibinfo {pages} {176602} (\bibinfo {year}
  {2004})%
  \bibAnnoteFile{NoStop}{Slot:04}%
\bibitem{Levy:06}%
  \BibitemOpen
  \bibfield{author}{%
  \bibinfo {author} {\bibfnamefont{E.}~\bibnamefont{Levy}}, \bibinfo {author}
  {\bibfnamefont{A.}~\bibnamefont{Tsukernik}}, \bibinfo {author}
  {\bibfnamefont{M.}~\bibnamefont{Karpovski}}, \bibinfo {author}
  {\bibfnamefont{A.}~\bibnamefont{Palevski}}, \bibinfo {author}
  {\bibfnamefont{B.}~\bibnamefont{Dwir}}, \bibinfo {author}
  {\bibfnamefont{E.}~\bibnamefont{Pelucchi}}, \bibinfo {author}
  {\bibfnamefont{A.}~\bibnamefont{Rudra}}, \bibinfo {author}
  {\bibfnamefont{E.}~\bibnamefont{Kapon}},\ and\ \bibinfo {author}
  {\bibfnamefont{Y.}~\bibnamefont{Oreg}},\ }%
  \bibfield{journal}{%
  \bibinfo {journal} {Phys. Rev. Lett.}\ }%
  \textbf{\bibinfo {volume} {97}},\ \bibinfo {pages} {196802} (\bibinfo {year}
  {2006})%
  \bibAnnoteFile{NoStop}{Levy:06}%
\bibitem{Kim:06}%
  \BibitemOpen
  \bibfield{author}{%
  \bibinfo {author} {\bibfnamefont{L.}~\bibnamefont{Venkataraman}}, \bibinfo
  {author} {\bibfnamefont{Y.~S.}\ \bibnamefont{Hong}},\ and\ \bibinfo {author}
  {\bibfnamefont{P.}~\bibnamefont{Kim}},\ }%
  \bibfield{journal}{%
  \Doi{10.1103/PhysRevLett.96.076601}{\bibinfo {journal} {Phys. Rev. Lett.}}\
  }%
  \textbf{\bibinfo {volume} {96}},\ \bibinfo {pages} {076601} (\bibinfo {year}
  {2006})%
  \bibAnnoteFile{NoStop}{Kim:06}%
\bibitem{KaneFis:92a}%
  \BibitemOpen
  \bibfield{author}{%
  \bibinfo {author} {\bibfnamefont{C.~L.}\ \bibnamefont{Kane}}\ and\ \bibinfo
  {author} {\bibfnamefont{M.~P.~A.}\ \bibnamefont{Fisher}},\ }%
  \bibfield{journal}{%
  \bibinfo {journal} {Phys. Rev. Lett.}\ }%
  \textbf{\bibinfo {volume} {68}},\ \bibinfo {pages} {1220} (\bibinfo {year}
  {1992})%
  \bibAnnoteFile{NoStop}{KaneFis:92a}%
\bibitem{KF:92b}%
  \BibitemOpen
  \bibfield{author}{%
  \bibinfo {author} {\bibfnamefont{C.~L.}\ \bibnamefont{Kane}}\ and\ \bibinfo
  {author} {\bibfnamefont{M.~P.~A.}\ \bibnamefont{Fisher}},\ }%
  \bibfield{journal}{%
  \Doi{10.1103/PhysRevB.46.15233}{\bibinfo {journal} {Phys. Rev. {\rm B}}}\ }%
  \textbf{\bibinfo {volume} {46}},\ \bibinfo {pages} {15233} (\bibinfo {year}
  {1992})%
  \bibAnnoteFile{NoStop}{KF:92b}%
\bibitem{MatYueGlaz:93}%
  \BibitemOpen
  \bibfield{author}{%
  \bibinfo {author} {\bibfnamefont{K.~A.}\ \bibnamefont{Matveev}}, \bibinfo
  {author} {\bibfnamefont{D.}~\bibnamefont{Yue}},\ and\ \bibinfo {author}
  {\bibfnamefont{L.~I.}\ \bibnamefont{Glazman}},\ }%
  \bibfield{journal}{%
  \bibinfo {journal} {Phys. Rev. Lett.}\ }%
  \textbf{\bibinfo {volume} {71}},\ \bibinfo {pages} {3351} (\bibinfo {year}
  {1993})%
  \bibAnnoteFile{NoStop}{MatYueGlaz:93}%
\bibitem{FurusakiNagaosa:93b}%
  \BibitemOpen
  \bibfield{author}{%
  \bibinfo {author} {\bibfnamefont{A.}~\bibnamefont{Furusaki}}\ and\ \bibinfo
  {author} {\bibfnamefont{N.}~\bibnamefont{Nagaosa}},\ }%
  \bibfield{journal}{%
  \bibinfo {journal} {Phys. Rev. B}\ }%
  \textbf{\bibinfo {volume} {47}},\ \bibinfo {pages} {4631} (\bibinfo {year}
  {1993})%
  \bibAnnoteFile{NoStop}{FurusakiNagaosa:93b}%
\bibitem{Furusaki:97}%
  \BibitemOpen
  \bibfield{author}{%
  \bibinfo {author} {\bibfnamefont{A.}~\bibnamefont{Furusaki}},\ }%
  \bibfield{journal}{%
  \Doi{10.1103/PhysRevB.56.9352}{\bibinfo {journal} {Phys. Rev. {\rm B}}}\ }%
  \textbf{\bibinfo {volume} {56}},\ \bibinfo {pages} {9352} (\bibinfo {year}
  {1997})%
  \bibAnnoteFile{NoStop}{Furusaki:97}%
\bibitem{EggertAffleck:95}%
  \BibitemOpen
  \bibfield{author}{%
  \bibinfo {author} {\bibfnamefont{S.}~\bibnamefont{Eggert}}\ and\ \bibinfo
  {author} {\bibfnamefont{I.}~\bibnamefont{Affleck}},\ }%
  \bibfield{journal}{%
  \bibinfo {journal} {Phys. Rev. Lett.}\ }%
  \textbf{\bibinfo {volume} {75}},\ \bibinfo {pages} {934} (\bibinfo {year}
  {1995})%
  \bibAnnoteFile{NoStop}{EggertAffleck:95}%
\bibitem{FabrizioGogolin:95}%
  \BibitemOpen
  \bibfield{author}{%
  \bibinfo {author} {\bibfnamefont{M.}~\bibnamefont{Fabrizio}}\ and\ \bibinfo
  {author} {\bibfnamefont{A.~O.}\ \bibnamefont{Gogolin}},\ }%
  \bibfield{journal}{%
  \bibinfo {journal} {Phys. Rev. B}\ }%
  \textbf{\bibinfo {volume} {51}},\ \bibinfo {pages} {17827} (\bibinfo {year}
  {1995})%
  \bibAnnoteFile{NoStop}{FabrizioGogolin:95}%
\bibitem{LL-Ph}%
  \BibitemOpen
  \bibfield{author}{%
  \bibinfo {author} {\bibfnamefont{P.}~\bibnamefont{San-Jose}}, \bibinfo
  {author} {\bibfnamefont{F.}~\bibnamefont{Guinea}},\ and\ \bibinfo {author}
  {\bibfnamefont{T.}~\bibnamefont{Martin}},\ }%
  \bibfield{journal}{%
  \bibinfo {journal} {Phys. Rev. {\rm B}}\ }%
  \textbf{\bibinfo {volume} {72}},\ \bibinfo {pages} {165427} (\bibinfo {year}
  {2005})%
  \bibAnnoteFile{NoStop}{LL-Ph}%
\bibitem{GYL:2011}%
  \BibitemOpen
  \bibfield{author}{%
  \bibinfo {author} {\bibfnamefont{A.}~\bibnamefont{Galda}}, \bibinfo {author}
  {\bibfnamefont{I.~V.}\ \bibnamefont{Yurkevich}},\ and\ \bibinfo {author}
  {\bibfnamefont{I.~V.}\ \bibnamefont{Lerner}},\ }%
  \bibfield{journal}{%
  \Doi{10.1103/PhysRevB.83.041106}{\bibinfo {journal} {Phys. Rev. B}}\ }%
  \textbf{\bibinfo {volume} {83}},\ \bibinfo {pages} {R041106} (\bibinfo {year}
  {2011})%
  \bibAnnoteFile{NoStop}{GYL:2011}%
\bibitem{GYL:2011a}%
  \BibitemOpen
  \bibfield{author}{%
  \bibinfo {author} {\bibfnamefont{A.}~\bibnamefont{Galda}}, \bibinfo {author}
  {\bibfnamefont{I.~V.}\ \bibnamefont{Yurkevich}},\ and\ \bibinfo {author}
  {\bibfnamefont{I.~V.}\ \bibnamefont{Lerner}},\ }%
  \bibfield{journal}{%
  \Doi{10.1209/0295-5075/93/17009}{\bibinfo {journal} {EPL}}\ }%
  \textbf{\bibinfo {volume} {93}},\ \bibinfo {pages} {17009} (\bibinfo {year}
  {2011})%
  \bibAnnoteFile{NoStop}{GYL:2011a}%
\bibitem{YuGYeL}%
  \BibitemOpen
  \bibfield{author}{%
  \bibinfo {author} {\bibfnamefont{I.~V.}\ \bibnamefont{Yurkevich}}, \bibinfo
  {author} {\bibfnamefont{A.}~\bibnamefont{Galda}}, \bibinfo {author}
  {\bibfnamefont{O.~M.}\ \bibnamefont{Yevtushenko}},\ and\ \bibinfo {author}
  {\bibfnamefont{I.~V.}\ \bibnamefont{Lerner}},\ }%
  \bibfield{journal}{%
  \bibinfo {journal} {Phys. Rev. Lett.}\ }%
  \textbf{\bibinfo {volume} {110}},\ \bibinfo {pages} {136405} (\bibinfo {year}
  {2013})%
  \bibAnnoteFile{NoStop}{YuGYeL}%
\bibitem{YuDual}%
  \BibitemOpen
  \bibfield{author}{%
  \bibinfo {author} {\bibfnamefont{I.~V.}\ \bibnamefont{Yurkevich}},\ }%
  \bibfield{journal}{%
  \bibinfo {journal} {Europhys. Lett.}\ }%
  \textbf{\bibinfo {volume} {104}},\ \bibinfo {pages} {37004} (\bibinfo {year}
  {2013})%
  \bibAnnoteFile{NoStop}{YuDual}%
\bibitem{Suppl-1}%
  \BibitemOpen
  \bibinfo {note} {The scattering state representation is described in Sect.~1
  in Supplemental Materials.}%
  \bibAnnoteFile{Stop}{Suppl-1}%
\bibitem{Suppl-2}%
  \BibitemOpen
  \bibinfo {note} {The explicit relation between $({\hat{\tilde R}}, \
  {\hat{\tilde T}})$ and $({\hat R}, \ {\hat T})$ is given in Sect.~2 in
  Supplemental Materials.}%
  \bibAnnoteFile{Stop}{Suppl-2}%
\bibitem{Suppl-3}%
  \BibitemOpen
  \bibinfo {note} {This step is described in Sect.~3 in Supplemental
  Materials.}%
  \bibAnnoteFile{Stop}{Suppl-3}%
\bibitem{Suppl-4}%
  \BibitemOpen
  \bibinfo {note} {The mathematical proof of this conclusion is given in
  Sect.~4 in Supplemental Materials.}%
  \bibAnnoteFile{Stop}{Suppl-4}%
\bibitem{AGR}%
  \BibitemOpen
  \bibfield{author}{%
  \bibinfo {author} {\bibfnamefont{A.}~\bibnamefont{Altland}}, \bibinfo
  {author} {\bibfnamefont{Y.}~\bibnamefont{Gefen}},\ and\ \bibinfo {author}
  {\bibfnamefont{B.}~\bibnamefont{Rosenow}},\ }%
  \bibfield{journal}{%
  \bibinfo {journal} {Phys. Rev. Lett.}\ }%
  \textbf{\bibinfo {volume} {108}},\ \bibinfo {pages} {136401} (\bibinfo {year}
  {2012})%
  \bibAnnoteFile{NoStop}{AGR}%
\bibitem{InhomLL}%
  \BibitemOpen
  \bibfield{author}{%
  \bibinfo {author} {\bibfnamefont{N.}~\bibnamefont{Sedlmayr}}, \bibinfo
  {author} {\bibfnamefont{J.}~\bibnamefont{Ohst}}, \bibinfo {author}
  {\bibfnamefont{I.}~\bibnamefont{Affleck}}, \bibinfo {author}
  {\bibfnamefont{J.}~\bibnamefont{Sirker}},\ and\ \bibinfo {author}
  {\bibfnamefont{S.}~\bibnamefont{Eggert}},\ }%
  \bibfield{journal}{%
  \bibinfo {journal} {Phys. Rev. {\rm B}}\ }%
  \textbf{\bibinfo {volume} {86}},\ \bibinfo {pages} {121302} (\bibinfo {year}
  {2012})%
  \bibAnnoteFile{NoStop}{InhomLL}%
\end{thebibliography}%


\onecolumngrid

\pagebreak

\section{Supplemental materials}

\subsection{1. Green's Functions}

Let us consider the Green's function, $ \, {\hat {\tilde{G}}}( x, x', \omega ) $, of non-interacting chiral bosons
\begin{equation}
  i {\hat{\tilde{ G}}} = \langle {\tilde{\boldsymbol\Theta}} \otimes {\tilde{\boldsymbol\Theta}}^{\rm T} \rangle \, ; \
    {\tilde{\boldsymbol  \Theta}}^{\rm T} = \left({\tilde{\boldsymbol  \theta}}_{\rm R}^{\rm T},{\tilde{\boldsymbol  \theta}}_{\rm L}^{\rm T}\right)\,,\,\,
    {\tilde{\boldsymbol  \theta}}^{\rm T}_{\eta}=\left( {\tilde \theta}_{\eta}^{(1)}, \ldots, {\tilde \theta}_{\eta}^{(N)}\right) \, .
\end{equation}
The Green's function without any scatterer is diagonal in the space of channels and its retarded component
reads [cf. Eq.(\ref{L0}) in the main text]:
\begin{equation}
   {\hat g}^{-1} = \frac{1}{2\pi}( {\hat \tau}_3 \partial_t + {\hat v} \partial_x) \partial_x \ \Rightarrow \
  {\hat g}(q; \omega_+) = \frac{2 \pi}{ ( {\hat \tau}_3 \omega_+ - {\hat v} q ) q} \, ; \quad \omega_+ \equiv \omega + i 0 \, ;
\end{equation}
with the Pauli matrices $ \, {\hat \tau}_k \, $ acting in the chiral space and the diagonal $ \, 2N \times 2 N \, $ matrix 
$ \, {\hat v} \, $ being defined in Eq.(\ref{V}) of the main text. Calculating the Fourier transform, we find:
\begin{equation}\label{Chiral-GF}
   {\hat g}(x - x'; \omega_+) = g_0 \, e^{ i \frac{|x-x'| \omega_+}{{\hat v}} }
       \left(
            \begin{array}{cc}
                \theta( x - x' ) & 0 \\
                                     0 & \theta( x' - x )
            \end{array}
       \right) - \frac{g_0}{2} {\hat \tau}_3 \ {\rm sign}(x-x') \, ; \quad g_0 \equiv - \frac{2 \pi i}{\omega_+} \, ;
\end{equation}
where $ \, \theta(x) \, $ is the step function.


Let us now add a scatterer at $ \, x = 0 $, which will be described by its transmission and reflection amplitudes. 
In the scattering state representation, we have to introduce incoming-/outgoing fields:
\begin{equation}
   {\tilde {\bf \Theta}}_{\rm out} \equiv
    \left(
     \begin{array}{c}
       {\tilde {\boldsymbol \theta}}_{\rm R}(x=+0) \\
       {\tilde {\boldsymbol \theta}}_{\rm L}(x=-0)
     \end{array}
   \right) \, ; \quad
   {\tilde {\bf \Theta}}_{\rm in} \equiv
    \left(
     \begin{array}{c}
       {\tilde {\boldsymbol \theta}}_{\rm L}(x=+0) \\
       {\tilde {\boldsymbol \theta}}_{\rm R}(x=-0)
     \end{array}
   \right) \, ;
\end{equation}
cf. Eq.(\ref{BoundCond-S}) in the main text. The correlation function of these fields
\begin{equation}
    i {\hat G}_{a,b} \equiv \langle {\tilde {\bf \Theta}}_{a} \otimes {\tilde {\bf \Theta}}_{b} \rangle \, , \quad
   a, b = {\rm in/out} \, ;
\end{equation}
can be introduced as usually:
\begin{enumerate}
  \item Correlations of ``in-in'' and ``out-out'' fields are not affected by the scatterer and, therefore:
       \begin{equation}\label{InIn}
            {\hat G}_{\rm in,in}(\omega) = {\hat G}_{\rm out,out}(\omega) = 
                                                             {\hat g} (\omega_+,0) = \frac{g_0}{2} \otimes 1_{2N \times 2N} \, ;
       \end{equation}
  \item Incoming fields are independent on the outgoing ones, i.e., they are not correlated:
       \begin{equation}\label{InOut}
            {\hat G}_{\rm in,out}(\omega) = 0 \, ;
       \end{equation}
  \item Correlations of outgoing fields with incoming ones are given by the scattering matrix:
       \begin{equation}\label{OutIn}
            {\hat G}_{\rm out,in}(\omega) = g_0 {\hat {\tilde S }} \, ; \quad
             {\hat {\tilde S }} = 
                \left(
                 \begin{array}{cc}
                   {\hat {\tilde R}} & {\hat {\tilde T}} \\
                   {\hat {\tilde T}} & {\hat {\tilde R}}
                 \end{array}
                \right);
       \end{equation}
    see Eq.(\ref{S-matr-Symm}) in the main text; here $ \, N \times N \, $ matrices $ \, \hat{T} \, $ and $ \, \hat{R} \, $ 
    are transmission and reflection amplitudes, respectively, and we have taken into account the symmetry of
    the scattering matrix with respect to the matrix $ \, {\hat \tau}_1 \, $ due to the reflection symmetry. 
\end{enumerate}


Combining Eqs.(\ref{InIn}--\ref{OutIn}), we determine the Green's function $ \, {\hat {\tilde G}} ( x, x' \to 0 ) $:
\begin{eqnarray}
    {\hat {\tilde G}}_{++} \equiv
    {\hat {\tilde G}}(x\to +0,x'\to +0) & = & g_0
                       \left(
                           \begin{array}{cc}
                                 1/2 & {\hat {\tilde R}} \\
                                    0 & 1/2 \\
                           \end{array}
                       \right) \, ; \\
    {\hat {\tilde G}}_{+-} \equiv
    {\hat {\tilde G}}(x\to +0,x'\to -0) & = & g_0
                       \left(
                           \begin{array}{cc}
                                 {\hat {\tilde T}} & 0 \\
                                                      0 & 0 \\
                           \end{array}
                       \right) \, ; \\
    {\hat {\tilde G}}_{-+} \equiv
    {\hat {\tilde G}}(x\to -0,x'\to +0) & = & g_0
                       \left(
                          \begin{array}{cc}
                             0 & 0 \\
                             0 & {\hat {\tilde T}} \\
                          \end{array}
                       \right) \, ; \\
    {\hat {\tilde G}}_{--} \equiv
   {\hat {\tilde G}}(x\to -0,x'\to -0) & = & g_0
                       \left(
                          \begin{array}{cc}
                               1/2 & 0 \\
                       {\hat {\tilde R}} & 1/2 \\
                          \end{array}
                       \right) \, .
\end{eqnarray}

The Green's function of the original chiral modes
\begin{equation}
  i {\hat G} = \langle {\boldsymbol\Theta} \otimes {\boldsymbol\Theta}^{\rm T} \rangle \, ; \
    {\boldsymbol \Theta}^{\rm T} = \left({\boldsymbol \theta}_{\rm R}^{\rm T},{\boldsymbol \theta}_{\rm L}^{\rm T}\right)\,,\,\,
    {\boldsymbol \theta}^{\rm T}_{\eta}=\left( \theta_{\eta}^{(1)}, \ldots, \theta_{\eta}^{(N)}\right) \, .
\end{equation}
is related to $ \, {\hat {\tilde G}} \, $ as
\begin{align}
{\hat G}(x,x')={\hat{\cal M}}\,{\hat {\tilde G}}(x,x')\,{\hat{\cal M}}^{\rm T} \, ,
\end{align}
see Eq.(\ref{RotGF}) in the main text. $ \, {\hat G}(x,x') \, $ is needed at $ \, x,x'=\pm 0 \, $ only, thus we introduce
$ \, {\hat G}_{\pm\pm}\equiv{\hat G}(x=\pm 0, x'=\pm 0) \, $ and express it in terms of $ \, {\hat {\tilde G}}_{\pm\pm} $:
\begin{align}\label{GF}
\left(
  \begin{array}{cc}
    {\hat G}_{++} & {\hat G}_{+-} \\
    {\hat G}_{-+} & {\hat G}_{--} \\
  \end{array}
\right)=
\left(
  \begin{array}{cc}
    {\cal M} {\hat {\tilde G}}_{++} {\cal M}^{\rm T} & {\cal M} {\hat {\tilde G}}_{+-} {\cal M}^{\rm T} \\
    {\cal M} {\hat {\tilde G}}_{-+} {\cal M}^{\rm T} & {\cal M} {\hat {\tilde G}}_{--} {\cal M}^{\rm T} \\
  \end{array}
\right)=g_0 \left(
          \begin{array}{cc}
            {\cal M}\left(
                      \begin{array}{cc}
                        1/2 & {\hat{\tilde R}} \\
                        0 & 1/2 \\
                      \end{array}
                    \right)
            {\cal M}^{\rm T} & {\cal M}\left(
                                 \begin{array}{cc}
                                   {\hat{\tilde T}} & 0 \\
                                   0 & 0 \\
                                 \end{array}
                               \right){\cal M}^{\rm T}
             \\
            {\cal M}\left(
                                 \begin{array}{cc}
                                  0 & 0 \\
                                   0 & {\hat{\tilde T}} \\
                                 \end{array}
                               \right){\cal M}^{\rm T} &  {\cal M}\left(
                      \begin{array}{cc}
                        1/2 & 0\\
                        {\hat{\tilde R}} & 1/2 \\
                      \end{array}
                    \right)
            {\cal M}^{\rm T} \\
          \end{array}
        \right) .
\end{align}

\subsection{2. Transfer matrix and scattering amplitudes}

Right $\leftrightarrow $ Left (inversion) symmetry of the system requires the following symmetry properties of
the scattering- and transfer- matrices:
\begin{align}
     {\hat{\cal S}}= {\hat \tau}_1\,{\hat{\cal S}}\,{\hat \tau}_1 \, , \quad
         {\rm and} \quad
    {\hat \tau}_1 \, {\hat{\cal T} } \, {\hat \tau}_1 =  {\hat{\cal T} }^{-1} \, .
\end{align}
The first symmetry relation means
\[
    {\hat{\cal S}}_{11} =  {\hat{\cal S}}_{22} \, , \quad  {\hat{\cal S}}_{12} =  {\hat{\cal S}}_{21} \, ,
\]
i.e. $ \, {\hat{\cal S}} \, $ can be parametrized with only two $N\times N$ reflection and transmission matrices:
\begin{equation}
     {\hat S} =
                       \left(
                           \begin{array}{cc}
                                   {\hat R} & {\hat T} \\
                                   {\hat T} & {\hat R} \\
                           \end{array}
                       \right) \, .
\end{equation}
The second symmetry relation results in
\begin{align}\label{T-restr}
\left(
  \begin{array}{cc}
    {\hat{\cal T}}_{22}{\hat{\cal T}}_{11}+{\hat{\cal T}}_{21}^2 &  {\hat{\cal T}}_{22}{\hat{\cal T}}_{12}+{\hat{\cal T}}_{21}{\hat{\cal T}}_{22} \\
    {\hat{\cal T}}_{11}{\hat{\cal T}}_{21}+{\hat{\cal T}}_{12}{\hat{\cal T}}_{11} & {\hat{\cal T}}_{11}{\hat{\cal T}}_{22}+{\hat{\cal T}}_{12}^2
  \end{array}
\right)=1 \, .
\end{align}
Thus, $ \,  {\hat{\cal T}} \, $ can be parametrized, for example, by elements $ \, {\hat{\cal T}}_{11} \, $ and $ \, {\hat{\cal T}}_{22} $:
\begin{align}
                        {\hat{\cal T}}=\left(
                                                  \begin{array}{cc}
                                                    {\hat{\cal T}}_{11} & \sqrt{1-{\hat{\cal T}}_{11}{\hat{\cal T}}_{22}} \\
                                                    -\sqrt{1-{\hat{\cal T}}_{22}{\hat{\cal T}}_{11}} & {\hat{\cal T}}_{22} \\
                                                  \end{array}
                                                \right) \, .
\end{align}

Using definitions of the scattering- and transfer- matrices, Eqs.(\ref{BoundCond-T},\ref{BoundCond-S}) in the main text,
and excluding $ \, {\hat{\cal T}}_{21} \, $ with the help of Eq.(\ref{T-restr}), one can find the relation between scattering and
transfer matrices:
\begin{align}
{\hat{\cal S}}=\left(
                 \begin{array}{cc}
                   {\hat{\cal T}}_{12}{\hat{\cal T}}_{22}^{-1} & \ {\hat{\cal T}}_{11}-{\hat{\cal T}}_{12}{\hat{\cal T}}_{22}^{-1}{\hat{\cal T}}_{21} \\
                   {\hat{\cal T}}_{22}^{-1} & -{\hat{\cal T}}_{22}^{-1}{\hat{\cal T}}_{21} \\
                 \end{array}
               \right)
=\left(
   \begin{array}{cc}
     {\hat{\cal T}}_{12}{\hat{\cal T}}_{22}^{-1} & {\hat{\cal T}}_{22}^{-1} \\
     {\hat{\cal T}}_{22}^{-1} & {\hat{\cal T}}_{12}{\hat{\cal T}}_{22}^{-1} \\
   \end{array}
 \right)=\left(
           \begin{array}{cc}
             {\hat R} & {\hat T} \\
             {\hat T} & {\hat R} \\
           \end{array}
         \right) .
\end{align}
Therefore, $ \, {\hat{\cal T}}_{22} = {\hat T}^{-1} \, $ and $ \, {\hat{\cal T}}_{12} = {\hat R} {\hat T}^{-1} $ and, using again Eq.(\ref{T-restr}),
we arrive at
\begin{align}
{\hat{\cal T}}=\left(
               \begin{array}{cc}
                 {\hat T}-{\hat R}{\hat T}^{-1}{\hat R} & {\hat R}{\hat T}^{-1} \\
                 -{\hat T}^{-1}{\hat R} & {\hat T}^{-1} \\
               \end{array}
             \right) .
\end{align}

%

After diagonalization of Lagrangian by the transformation
\begin{equation}
    {\boldsymbol\Theta}(x) = {\hat{\cal M}}\,{\tilde{\boldsymbol\Theta}}(x) \, ,
\end{equation}
we have to rotate the transfer-matrix
\begin{align}
   {\hat{\tilde{\cal T}}}={\hat{\cal M}}^{-1}\,{\hat{\cal T}}\,{\hat{\cal M}} \, .
\end{align}
Note that $ \, {\hat{\cal M}}=\tau_1\,{\hat{\cal M}}\,\tau_1 \, $ and it can be parametrized as follows
\begin{equation}\label{M-L-param}
    {\hat{\cal M}} = {\hat L}^{-1} \, {\hat M}  {\hat L} \, ; \qquad
    {\hat M} = {\rm diag} \{ {\hat M}_1 \, , \ {\hat M}_2\} \, ;
\end{equation}
where
\begin{equation}
    {\hat L} = \frac{1}{\sqrt{2}}
                \left(
                  \begin{array}{cc}
                               1   & -1 \\
                               1   &  1
                  \end{array}
                \right) \, , \quad
    {\hat L}^{-1} = {\hat L}^{\rm T} \, .
\end{equation}
Eq.(\ref{M-L-param}) is equivalent to Eq.(\ref{M-parametr}) in the main text after substituting $ \,   {\hat M} \to {\hat M}_1 \, ,
\mbox{ and } \left( {\hat M}^{\rm -1} \right)^{\rm T}\to {\hat M}_2 $.
The symmetry of the transfer matrix in the system with the inversion symmetry ensures that the
matrix $ \, {\hat{\tilde {\cal T}}} \, $ has the same structure as $ \, {\hat{\cal T}} $, but with different
scattering amplitudes:
\begin{align}
{\hat{\tilde {\cal T}}}=\left(
               \begin{array}{cc}
                 {\hat {\tilde T}}-{\hat {\tilde R}}{\hat {\tilde T}}^{-1}{\hat {\tilde R}} & \ {\hat {\tilde R}}{\hat {\tilde T}}^{-1} \\
                 -{\hat {\tilde T}}^{-1}{\hat {\tilde R}} & {\hat {\tilde T}}^{-1} \\
               \end{array}
             \right) .
\end{align}
Now we will find relations between $ \, ( {\hat {\tilde R}}, \, {\hat {\tilde T}} ) \, $ and $ \, ( {\hat R}, \, {\hat T} ) $.

Let us introduce (auxiliary) rotated transfer matrices:
\begin{align}
{\hat{{\cal T}}}^{L}={\hat L}\,{\hat{{\cal T}}}\,{\hat L}^{-1}=\frac{1}{2}
\left(
  \begin{array}{cc}
    {\hat T}+(1-{\hat R})\,{\hat T}^{-1}\,(1+{\hat R}) & \ {\hat T}-(1-{\hat R})\,{\hat T}^{-1}\,(1-{\hat R}) \\
    {\hat T}-(1+{\hat R})\,{\hat T}^{-1}\,(1+{\hat R}) & \ {\hat T}+(1+{\hat R})\,{\hat T}^{-1}\,(1-{\hat R}) \\
  \end{array}
\right) \, ;
\end{align}
and
\begin{align}\label{T-L}
{\hat{{\tilde {\cal T}}}}^{L}={\hat L}\,{\hat{{\tilde{\cal T}}}}\,{\hat L}^{-1}=\frac{1}{2}
\left(
  \begin{array}{cc}
    {\hat {\tilde T}}+(1-{\hat {\tilde R}})\,{\hat {\tilde T}}^{-1}\,(1+{\hat {\tilde R}}) &
                                                \ {\hat {\tilde T}}-(1-{\hat {\tilde R}})\,{\hat {\tilde T}}^{-1}\,(1-{\hat {\tilde R}}) \\
    {\hat {\tilde T}}-(1+{\hat {\tilde R}})\,{\hat {\tilde T}}^{-1}\,(1+{\hat {\tilde R}}) &
                                                \ {\hat {\tilde T}}+(1+{\hat {\tilde R}})\,{\hat {\tilde T}}^{-1}\,(1-{\hat {\tilde R}}) \\
  \end{array}
\right) \, .
\end{align}
The matrix $ \, {\hat{{\tilde {\cal T}}}}^{L} \, $ can also be expressed via entries of $ \, {\hat{{\cal T}}}^{L} $:
\begin{equation}\label{T-L-Tilde}
   {\hat{\tilde{\cal T}}}^{L} =  {\hat M}^{-1} \, {\hat{{\cal T}}}^{L}  \, {\hat M} =
                                          \left(
                                           \begin{array}{cc}
                                             {\hat M}^{-1}_1 {\hat{\cal T}}^{L}_{11} {\hat M}_1 & {\hat M}^{-1}_1 {\hat{\cal T}}^{L}_{12} {\hat M}_2 \\
                                             {\hat M}^{-1}_2 {\hat{\cal T}}^{L}_{21} {\hat M}_1 & {\hat M}^{-1}_2 {\hat{\cal T}}^{L}_{22} {\hat M}_2
                                           \end{array}
                                          \right) \, .
\end{equation}
Using the identity $ \, {\hat L}^{-1} {\hat \tau}_1  {\hat L} = {\hat \tau}_3 $, one can prove that the matrix $ \, \hat{{\cal T}}^{L} \, $
obeys the symmetry
\begin{align}
  {\hat \tau}_3 \, {\hat{\cal T}}^{L} \, {\hat \tau}_3 = \left( {\hat{\cal T}}^{L} \right)^{-1} \, ,
\end{align}
which means
\begin{align}
\left(
  \begin{array}{cc}
    \left[{\hat{\cal T}}^{L}_{11}\right]^2-{\hat{\cal T}}^{L}_{12}{\hat{\cal T}}^{L}_{21} & \ -{\hat{\cal T}}^{L}_{11}{\hat{\cal T}}^{L}_{12}+{\hat{\cal T}}^{L}_{12}{\hat{\cal T}}^{L}_{22} \\
    {\hat{\cal T}}^{L}_{21}{\hat{\cal T}}^{L}_{11}-{\hat{\cal T}}^{L}_{22}{\hat{\cal T}}^{L}_{21} & \ \left[{\hat{\cal T}}^{L}_{22}\right]^2-{\hat{\cal T}}^{L}_{21}{\hat{\cal T}}^{L}_{12} \\
  \end{array}
\right)=1 \, .
\end{align}
Thus, we can parametrize $ \, {\hat{\cal T}}^{L} \, $ by, for example, its off-diagonal entries:
\begin{align}
   {\hat{\cal T}}^{L}_{11} = \sqrt{1+{\hat{\cal T}}^{L}_{12}{\hat{\cal T}}^{L}_{21}} \, , \quad
   {\hat{\cal T}}^{L}_{22} = \sqrt{1+{\hat{\cal T}}^{L}_{21}{\hat{\cal T}}^{L}_{12}} \, .
\end{align}
A straightforward algebra yields:
\begin{eqnarray}\label{T-L-mod}
{\hat{\cal T}}^{L} & = & \frac{1}{2}
\left(
  \begin{array}{cc}
    {\hat{\cal T}}^{L}_{11}  & -(1-{\hat R}+{\hat T})\,{\hat T}^{-1}\,(1-{\hat R}-{\hat T}) \\
    -(1+{\hat R}+{\hat T})\,{\hat T}^{-1}\,(1+{\hat R}-{\hat T}) &  {\hat{\cal T}}^{L}_{22} \\
  \end{array}
\right) = \\
                            & = &
\left(
  \begin{array}{cc}
    {\hat{\cal T}}^{L}_{11} & \left[(1-{\hat R}+{\hat T})^{-1}-(1-{\hat R}-{\hat T})^{-1}\right]^{-1} \\
    \left[(1+{\hat R}+{\hat T})^{-1}-(1+{\hat R}-{\hat T})^{-1}\right]^{-1} & {\hat{\cal T}}^{L}_{22} \\
  \end{array}
\right) \, .
\end{eqnarray}
The same manipulations with Eq.(\ref{T-L-Tilde}) yield:
\begin{eqnarray}\label{T-L-Tilde-mod}
{\hat {\tilde {\cal T}}}^{L} =
\left(
  \begin{array}{cc}
    {\hat{\tilde {\cal T}}}^{L}_{11} & \left[(1-{\hat {\tilde R}}+{\hat {\tilde T}})^{-1}-(1-{\hat {\tilde R}}-{\hat {\tilde T}})^{-1}\right]^{-1} \\
    \left[(1+{\hat {\tilde R}}+{\hat {\tilde T}})^{-1}-(1+{\hat {\tilde R}}-{\hat {\tilde T}})^{-1}\right]^{-1} & {\hat{\tilde {\cal T}}}^{L}_{22} \\
  \end{array}
\right) \, .
\end{eqnarray}


Combining Eqs.(\ref{T-L-Tilde},\ref{T-L-mod},\ref{T-L-Tilde-mod}) and inverting matrix entries, we find:
\begin{align}
\label{Eq-12}
\mbox{from entries \{1,2\}:} \
\left(1-{\hat {\tilde R}}+{\hat {\tilde T}}\right)^{-1}-\left(1-{\hat {\tilde R}}-{\hat {\tilde T}}\right)^{-1}&={\hat M}_2^{-1}\,\left[\left(1-{\hat R}+{\hat T}\right)^{-1}-\left(1-{\hat R}-{\hat T}\right)^{-1}\right]\,{\hat M}_1 \, ; \\
\label{Eq-21}
\mbox{from entries \{2,1\}:} \
\left(1+{\hat {\tilde R}}+{\hat {\tilde T}}\right)^{-1}-\left(1+{\hat {\tilde R}}-{\hat {\tilde T}}\right)^{-1}&={\hat M}_1^{-1}\,\left[\left(1+{\hat R}+{\hat T}\right)^{-1}-\left(1+{\hat R}-{\hat T}\right)^{-1}\right]\,{\hat M}_2 \, .
\end{align}
To solve these equations we parametrize
\begin{align}
   {\hat S}_{\pm}={\hat R}\pm{\hat T}=\frac{1-{\hat\xi}_{\pm}}{1+{\hat\xi}_{\pm}} \, ,
                     \quad \Rightarrow \quad
   1-{\hat S}_{\pm}=\frac{2{\hat\xi}_{\pm}}{1+{\hat\xi}_{\pm}} \,\mbox{ and } \,
   1+{\hat S}_{\pm}=\frac{2}{1+{\hat\xi}_{\pm}} \, ;
\end{align}
and
\begin{align}\label{Stild-Xitild}
   {\hat {\tilde S}}_{\pm}={\hat {\tilde R}}\pm{\hat {\tilde T}}=\frac{1-{\hat{\tilde \xi}}_{\pm}}{1+{\hat{\tilde \xi}}_{\pm}} \, ,
                     \quad \Rightarrow \quad
   1-{\hat {\tilde S}}_{\pm}=\frac{2{\hat{\tilde \xi}}_{\pm}}{1+{\hat{\tilde \xi}}_{\pm}} \,\mbox{ and } \,
   1+{\hat {\tilde S}}_{\pm}=\frac{2}{1+{\hat{\tilde \xi}}_{\pm}} \, ;
\end{align}
and reduce Eqs.(\ref{Eq-12},\ref{Eq-21}) to:
\begin{align}
   \label{xi-pm-1}
{\hat{\tilde\xi}}_{+}^{-1}-{\hat{\tilde\xi}}_{-}^{-1}&={\hat M}_2^{-1}\,\left[{\hat\xi}_{+}^{-1}-{\hat\xi}_{-}^{-1}\right]\,{\hat M}_1 \, , \\
   \label{xi-pm-2}
{\hat{\tilde\xi}}_{+}-{\hat{\tilde\xi}}_{-}&={\hat M}_1^{-1}\left[{\hat\xi}_{+}-{\hat\xi}_{-}\right]\,{\hat M}_2 \, .
\end{align}
The obvious solution of Eqs.(\ref{xi-pm-1},\ref{xi-pm-2}), which relates $ \, ( {\hat {\tilde R}}, \, {\hat {\tilde T}} ) \, $ and
$ \, ( {\hat R}, \, {\hat T} ) $, reads:
\begin{align}\label{xi-pm-sol1}
{\hat{\tilde\xi}}_{\pm}= {\hat M}_1^{-1}\,{\hat\xi}_{\pm}\,{\hat M}_2 \, .
\end{align}
The second solution of Eqs.(\ref{xi-pm-1},\ref{xi-pm-2}) can be obtained after noticing that they
are invariant with respect to transformation $ \, {\hat{\tilde\xi}}_{+} \leftrightarrow - {\hat{\tilde\xi}}_{-} $.
Applying this transformation to Eq.(\ref{xi-pm-sol1}) we find:
\begin{align}\label{xi-pm-sol2}
{\hat{\tilde\xi}}_{\pm}= - {\hat M}_1^{-1}\,{\hat\xi}_{\mp}\,{\hat M}_2 \, .
\end{align}
Note, however, that the solutions Eq.(\ref{xi-pm-sol2}) does not satisfy the continuity. Namely,
in the non-interacting case, where $ \, {\hat M}_1 = {\hat M}_2 $, we require $ \, {\hat{\tilde\xi}}_{\pm} =
{\hat\xi}_{\pm} $ which holds true only for the solution Eq.(\ref{xi-pm-sol1}).


\subsection{3. Auto-correlation function $ \, {\hat \Delta} $}

For both weak scatterer and weak link perturbations can be written in a unified way,
see Eqs.(\ref{pert-1},\ref{pert-2}) in the main text:
\begin{eqnarray}
   L_{\rm ws} & = & \lambda \cos\Bigl( \Phi(T=1) \Bigr) \, ; \\
   L_{\rm wl} & = & t \cos\Bigl( \Phi(T=0) \Bigr) \, ;
\end{eqnarray}
where the field $ \, \Phi(T) \, $ is the difference between two incoming chiral fields of the 1st
channel for the conducting phase ($ T = 1$) and for the insulating phase ($ T = 0 $):
\begin{equation}
   \label{Phi-cp-ip}
   \Phi(T) = \left[ \theta^{(1)}_{\rm R}(x=-0,t) - \theta^{(1)}_{\rm L}(x=+0,t) \right] \
                        \mbox{ at } \ 
   \left\{
    \begin{array}{l}
      T = 1 \, , \ R = 0  \quad \mbox{for CP} \\
      T = 0 \, , \ R = 1  \quad \mbox{for IP}
    \end{array}
   \right.
\end{equation}
%
The notation in Eq. (\ref{Phi-cp-ip}) stresses that the field $ \, \Phi(T) \, $
and its auto-correlation function depend on boundary conditions including, of course, boundary conditions
in the LL-channel. In particular, one can restore formulas for the same perturbations introduced in Ref.\cite{YuGYeL}
by using the matching conditions
\begin{eqnarray}
   \theta^{(1)}_{\rm R,L}(x=-0,t) & = & \theta^{(1)}_{\rm R,L}(x=+0,t) \ \mbox{ in the CP} \, ; \\
   \theta^{(1)}_{\rm R}(x=\pm 0, t) & = & \theta^{(1)}_{\rm L}(x=\pm 0, t) \ \mbox{ in the IP} \, .
\end{eqnarray}
As discussed in the main text, the RG equations are governed by the retarded component of the Green's function 
$ \, {\cal G}(t-t';T) = - i \left\langle  \, \Phi(t;T) \, \Phi(t';T) \, \right\rangle $:
\begin{equation}
   {\cal G}(\omega_+;T) \equiv g_0 \, \left[ \hat{\Delta}(T) \right]_{11} \, .
\end{equation}
Here we have defined the $ \, N \times N \, $ matrix $ \, \hat{\Delta}(T) \, $ which depends on the phase
(either the CP or the IP) and
can be found by using the Green's functions introduced in the Sect.1 of the Supplemental materials:
\begin{equation}
  {\hat \Delta} = \frac{1}{g_0} \left[{\hat G}^{RR}_{--}+{\hat G}^{LL}_{++}-{\hat G}^{RL}_{-+}-{\hat G}^{LR}_{+-}\right] \, ;
\end{equation}
superscripts are related to chirality indices. Using Eq.(\ref{GF}) we arrive at
\begin{align}\label{Delta-Generic}
{\hat \Delta}=\frac{1}{2} {\rm tr}_{\rm ch}\,{\hat{\cal M}}\left(
                                                \begin{array}{cc}
                                                  1 & {\hat{\tilde S}}_- \\
                                                  {\hat{\tilde S}}_- & 1 \\
                                                \end{array}
                                              \right)
 \,{\hat {\cal M}}^{\rm T}\,,\quad {\hat{\tilde S}}_-={\hat {\tilde R}}-{\hat {\tilde T}} \, .
\end{align}
Trace $ \, {\rm tr}_{\rm ch} \, $ is calculated over the chiral space of right-/left- movers; expanding it in Eq.(\ref{Delta-Generic})
we find:
\begin{equation}\label{Delta-Expanded}
   {\hat \Delta}=\frac{1}{2}
         \left[
            \hat{M}_1 \left( 1 - {\hat {\tilde S}}_ {-} \right) \hat{M}_1^{\rm T} +
            \hat{M}_2 \left( 1 + {\hat {\tilde S}}_ {-} \right) \hat{M}_2^{\rm T}
         \right] \, .
\end{equation}
Now we a) use Eqs.(\ref{Stild-Xitild},\ref{xi-pm-sol1}) to express $ \, {\hat {\tilde S}}_ {-} \, $ in terms of $ \, {\hat \xi}_{-} \, $
and matrices $ \, \hat{M}_{1,2} $; and b) simplify Eq.(\ref{Delta-Expanded}) to
\begin{align}\label{Delta-Final}
{\hat\Delta} = \left[{\hat\xi}_-^{-1}+{\hat\delta}^{-1}\right]^{-1}\,{\hat M}_2{\hat M}_1^{\rm T} +
                       \left[{\hat\xi}_-+{\hat\delta}\right]^{-1}\,\,{\hat M}_1{\hat M}_2^{\rm T} \, ;
             \quad
                      {\hat\delta} \equiv {\hat M}_1 \, {\hat M}_2^{-1} \, .
\end{align}
The symmetry Eq.(\ref{Symmetry1}) in the main text implies
\begin{equation}
{\hat M}_1{\hat M}_2^{\rm T}= 1 \, .
\end{equation}
Taking this into account (and skipping subscripts of  $ \, {\hat \xi}_-  \, $ and $ \, {\hat M}_1 $),  we reduce
Eqs.(\ref{Delta-Expanded},\ref{Delta-Final}) to Eqs.(\ref{HatDelta},\ref{Delta-answ}) of the main text.

\subsection{4. Duality of Matrix Elements}

Let us consider $ \, N \times N \, $ symmetric matrix and write it  as a block matrix
\begin{equation}\label{BlDecomp}
   \left(
     \begin{array}{cc}
           a_0 & \psi^{\rm T} \\
           \psi  & {\hat b} \\
     \end{array}
   \right)
\end{equation}
where $ \, a_0 \, $ is scalar, $\psi$ is $ \, (N-1) $-dimensional vector  and ${\hat b}$ is $ (N-1) \times (N-1) $ symmetric matrix .
Its inversion is given by
\begin{align}\label{inv}
\left(
  \begin{array}{cc}
    a_0 & \psi^{\rm T} \\
    \psi & {\hat b} \\
  \end{array}
\right)^{-1}=\left(
               \begin{array}{cc}
                 A & -A\left[{\hat b}^{-1}\psi\right]^{\rm T} \\
                 -A{\hat b}^{-1}\psi & \left[ {\hat b}-\psi\otimes\psi^{\rm T}/a_0 \right]^{-1} \\
               \end{array}
             \right)\,,\qquad A= \left( a_0 - \psi^{\rm T}{\hat b}^{-1}\psi \right)^{-1} \, .
\end{align}

Our goal is to find matrix entries
\begin{equation}
   \Delta_{\rm ws} = \lim_{\xi \to \infty} \left[ \frac{1}{{\hat\delta}^{-1} + {\hat \xi}^{-1}} \right]_{11} \, , \quad
   \Delta_{\rm wl} = \lim_{\xi \to 0}
                                  \left[ \frac{1}{{\hat\delta}+{\hat \xi}} \right]_{11} \, ;
\end{equation}
where $ \, \xi = {\hat \xi}_{11} $, $ \, {\hat \delta} = \hat{M} \hat{M}^{\rm T} $, and the matrices $ \, {\hat\delta} \, $ and
$ \, {\hat \xi} \, $  can be written in the decomposition (\ref{BlDecomp}) as follows:
\begin{align}
{\hat\delta} =
                    \left(
                      \begin{array}{cc}
                        \delta & \psi^{\rm T}\ \\
                        \psi & {\hat\delta}_{\rm e} \\
                      \end{array}
                    \right) , \quad
{\hat\xi}    =\left(
                      \begin{array}{cc}
                            \xi & 0 \\
                            0   & {\hat \zeta} \\
                      \end{array}
                    \right) ;
\end{align}
see. Eqs.(\ref{Delta-answ},\ref{xi}) in the main text.


Using Eq.(\ref{inv}), we find
\begin{align}\label{wl}
\Delta_{\rm wl}=\left(
                      \begin{array}{cc}
                        \delta & \psi^{\rm T}\ \\
                        \psi & {\hat\delta}_{\rm e}+{\hat\zeta} \\
                      \end{array}
                    \right)^{-1}_{11}=\left[\delta-\psi^{\rm T}\left({\hat\delta}_{\rm e}+{\hat\zeta}\right)^{-1}\psi\right]^{-1} \, .
\end{align}
To calculate $ \, \Delta_{\rm ws} $, let us use the identity
\begin{align}
\left({\hat \alpha}^{-1}+{\hat \beta}^{-1}\right)^{-1}={\hat \alpha}\left({\hat \alpha}+{\hat \beta}\right)^{-1}{\hat \beta}
\end{align}
which holds true for non-singular matrices $ \, {\hat \alpha} \, $ and $ \, {\hat \beta} $ and allows us to write the formula
for $ \, \Delta_{\rm ws} \, $ as follows:

\begin{align}\label{DwlLim}
\Delta_{\rm ws}=\lim_{\xi\to\infty}\left[{\hat\delta}\left({\hat\delta}+{\hat \xi}\right)^{-1} {\hat \xi} \right]_{11} =
                           \left[{\hat\delta} \lim_{\xi\to\infty} \left\{ \xi \left({\hat\delta}+{\hat \xi}\right)^{-1} \right\} \right]_{11}
\end{align}
The limit $ \, \xi \to \infty \, $ is easily found from the inversion Eq. (\ref{inv})

\begin{align}\label{Unc}
\lim_{\xi\to\infty}\left\{ \xi \left({\hat\delta}+{\hat \xi}\right)^{-1} \right\}=
                                               \left(
                                                     \begin{array}{cc}
                                                            1 & -\left[\left({\hat\delta}_{\rm e}+{\hat\zeta}\right)^{-1}\psi\right]^{\rm T} \\
                                                           -\left({\hat\delta}_{\rm e}+{\hat\zeta}\right)^{-1}\psi & - \psi \otimes \psi^{\rm T} \\
                                                     \end{array}
                                               \right) \, .
\end{align}
Inserting Eq.(\ref{Unc}) into Eq.(\ref{DwlLim}), we find
\begin{align}\label{ws}
   \Delta_{\rm ws}=\delta-\psi^{\rm T}\left({\hat\delta}_{\rm e}+{\hat\zeta}\right)^{-1}\psi \quad \Rightarrow \quad
 \Delta_{\rm ws} =\Delta_{\rm wl}^{-1} \, .
\end{align}

\end{document}